\documentstyle[epsfig]{aipproc}

\begin{document}

$\null$
hep-ph/9706392 \hfill UCLA/97/TEP/13
\title{ Two-Loop N=4 Supersymmetric Amplitudes and QCD\thanks{This work was
supported by the DOE under contract DE-FG03-91ER40662 and by the
Alfred P. Sloan Foundation under grant BR-3222.}}

\author{Zvi Bern\thanks{Presenter at 5th International Workshop on 
Deep Inelastic Scattering and QCD, April, 1997}, 
Joel Rozowsky and Bryce Yan}
\address{Department of Physics, UCLA, Los Angeles, CA 90095}

\maketitle

\vskip -.5 cm 

\begin{abstract}
Two-loop four-gluon $N=4$ susy amplitudes are evaluated via cutting
techniques as a testing ground for QCD. A conjecture for four-point
amplitudes to all loop orders is described. We also present a new
conjecture for the leading-color part of the two-loop five-gluon
amplitudes.
\end{abstract}

\def\fig#1{Fig.~{\ref{#1}}}
\def\eqn#1{Eq.~(\ref{#1})}
\def\tree{{\rm tree}}

\newskip\humongous \humongous=0pt plus 1000pt minus 1000pt
\def\caja{\mathsurround=0pt}
\def\eqalign#1{\,\vcenter{\openup1\jot \caja
        \ialign{\strut \hfil$\displaystyle{##}$&$
        \displaystyle{{}##}$\hfil\crcr#1\crcr}}\,}
\newif\ifdtup
\def\panorama{\global\dtuptrue \openup1\jot \caja
        \everycr{\noalign{\ifdtup \global\dtupfalse
        \vskip-\lineskiplimit \vskip\normallineskiplimit
        \else \penalty\interdisplaylinepenalty \fi}}}
\def\eqalignno#1{\panorama \tabskip=\humongous
        \halign to\displaywidth{\hfil$\displaystyle{##}$
        \tabskip=0pt&$\displaystyle{{}##}$\hfil
        \tabskip=\humongous&\llap{$##$}\tabskip=0pt
        \crcr#1\crcr}}

%
\newcounter{eqnumber}
\renewcommand{\theeqnumber}{\arabic{eqnumber}}
\def\equn{
\refstepcounter{eqnumber}
\eqno({\rm \theeqnumber})
}


\def\tr{{\rm tr}}

\newbox\charbox
\newbox\slabox
\def\s#1{{      
        \setbox\charbox=\hbox{$#1$}
        \setbox\slabox=\hbox{$/$}
        \dimen\charbox=\ht\slabox
        \advance\dimen\charbox by -\dp\slabox
        \advance\dimen\charbox by -\ht\charbox
        \advance\dimen\charbox by \dp\charbox
        \divide\dimen\charbox by 2
        \raise-\dimen\charbox\hbox to \wd\charbox{\hss/\hss}
        \llap{$#1$}
}}

\def\spa#1.#2{\left\langle#1\,#2\right\rangle}
\def\spb#1.#2{\left[#1\,#2\right]}
\def\lor#1.#2{\left(#1\,#2\right)}
\def\sand#1.#2.#3{%
  \left\langle\smash{#1}{\vphantom1}\right|{#2}%
  \left|\smash{#3}{\vphantom1}\right\rangle}
\def\sandp#1.#2.#3{%
  \left\langle\smash{#1}{\vphantom1}^{-}\right|{#2}%
  \left|\smash{#3}{\vphantom1}^{+}\right\rangle}
\def\sandpp#1.#2.#3{%
  \left\langle\smash{#1}{\vphantom1}^{+}\right|{#2}%
  \left|\smash{#3}{\vphantom1}^{+}\right\rangle}
\def\sandmm#1.#2.#3{%
  \left\langle\smash{#1}{\vphantom1}^{-}\right|{#2}%
  \left|\smash{#3}{\vphantom1}^{-}\right\rangle}
\def\sandpm#1.#2.#3{%
  \left\langle\smash{#1}{\vphantom1}^{+}\right|{#2}%
  \left|\smash{#3}{\vphantom1}^{-}\right\rangle}
\def\sandmp#1.#2.#3{%
  \left\langle\smash{#1}{\vphantom1}^{-}\right|{#2}%
  \left|\smash{#3}{\vphantom1}^{+}\right\rangle}
\catcode`@=11  

\def\si{\sigma}
\def\Tr{\, {\rm Tr}}
\def\P{{\rm P}}
\def\NP{{\rm NP}}
\def\LC{{\rm LC}}
\def\SC{{\rm SC}}
\def\eps{\epsilon}
\def\Ord{{\cal O}}
\def\neqfour{N=4}
 \def\scut{s\ {\rm cut}}
\def\I{{\cal I}}
\def\fourperm#1#2#3#4{(#1\,#2\,#3\,#4)}%
\def\susy{{\scriptscriptstyle \rm SUSY}}
\def\pol{\eps}


\section*{Introduction}

Over the years there have been a number of rather impressive Feynman
diagram calculations at two and higher loops. (For one recent example
see ref.~\cite{FourLoopBeta}).  However, a number of important
computations remain to be performed.  Two examples of computations
which are required for analysis of experiments but have not been
carried out are two-loop contributions to $e^+ \, e^-
\rightarrow 3$ jets and to Altarelli-Parisi splitting functions.  More
generally, no computations have appeared at two and higher loops which
involve more than a single kinemtic variable.

In contrast, at one-loop, recent years have seen 
improvements in the calculational techniques for scattering amplitudes
\cite{Review}.  In this talk we discuss initial steps taken to apply
some of these techniques to two- and higher-loop amplitudes.  In
particular, we apply the recent developments in cutting techniques for
obtaining complete amplitudes with no ambiguities or subtractions.
These cutting techniques have been used for obtaining infinite
sequences of one-loop supersymmetric amplitudes
\cite{InfiniteSequences} and for the calculations performed in
refs.~\cite{FourPartons} of the one-loop helicity amplitudes for
$e^+\, e^- \rightarrow 4$ jets and deeply inelastic scattering in
association with three jets.  Other approaches for efficiently
calculating higher-loop amplitudes taken by various authors include
string theory \cite{Magnea}, first quantized \cite{FirstQuantized} and
recursive \cite{Nair} approaches. (Further references are contained in
ref.~\cite{Review}.)

A good starting point for exploring cutting techniques for
higher-loops are $N=4$ supersymmetric amplitudes since they are
particularly simple.  Using cutting techniques we have obtained the
exact expressions for the two-loop four-gluon $N=4$ amplitudes
\cite{TwoLoopSusyFour}.
Furthermore, with cutting techniques we have obtained conjectures for
the structure of two-loop five-gluon amplitudes and for four-point
amplitudes to arbitrary loop orders. We expect a study of $N=4$
amplitudes to be useful for obtaining two- and higher-loop QCD
amplitudes via cutting techniques.

\section*{Properties of N=4 Supersymmetric Amplitudes}

In performing the calculation we apply some of the techniques
that have proven useful at tree level and at one loop.  These include
spinor helicity, color decompositions and supersymmetry identities.
For references and further details the reader may consult various 
reviews \cite{ManganoReview,Review}.

The high degree of supersymmetry present in $N\! =\! 4$ amplitudes
considerably simplifies their analytic structure.  At one-loop the
leading color $N\! =\! 4$ partial amplitudes are rather simple and
are given by \cite{GSB}
$$
A_{4;1}^{1 \mbox{-} \rm loop}(1,2,3,4) = i st \, A_4^{\tree}(1,2,3,4)
 \, \I_4^{1 \mbox{-} \rm loop}(s,t) \, ,
\equn\label{OneLoopNFour}
$$
where $ \I_4^{1 \mbox{-} \rm loop}(s,t)$ is the one-loop scalar
integral and the tree amplitudes may be found in
ref.~\cite{ManganoReview}. The Mandelstam variables are defined as
$s\equiv (k_1+k_2)^2$ and $t\equiv (k_2+k_3)^2$.  The massless $N \, =
\, 4$ super-multiplet consists of one gluon, four Weyl fermions and
six real scalars, whose contributions are summed over to obtain the
result (\ref{OneLoopNFour}).  The subleading color contributions may
be obtained in terms of sums of permutations of the expression
(\ref{OneLoopNFour}).

When using cuts to obtain the two-loop $N=4$ gluon amplitudes we also
need tree and one-loop amplitudes with either two external scalars or
fermions, since these particles can also cross the cuts. The maximally
helicity violating amplitudes with two external fermions or scalars
may be obtained directly from the gluon amplitudes using supersymmetry
identities.

\section*{Construction of Two-Loop Amplitudes}
\label{TwoLoopCutSection}

%
\begin{figure}[ht]
\begin{center}
\vskip -.2 cm 
\epsfig{file=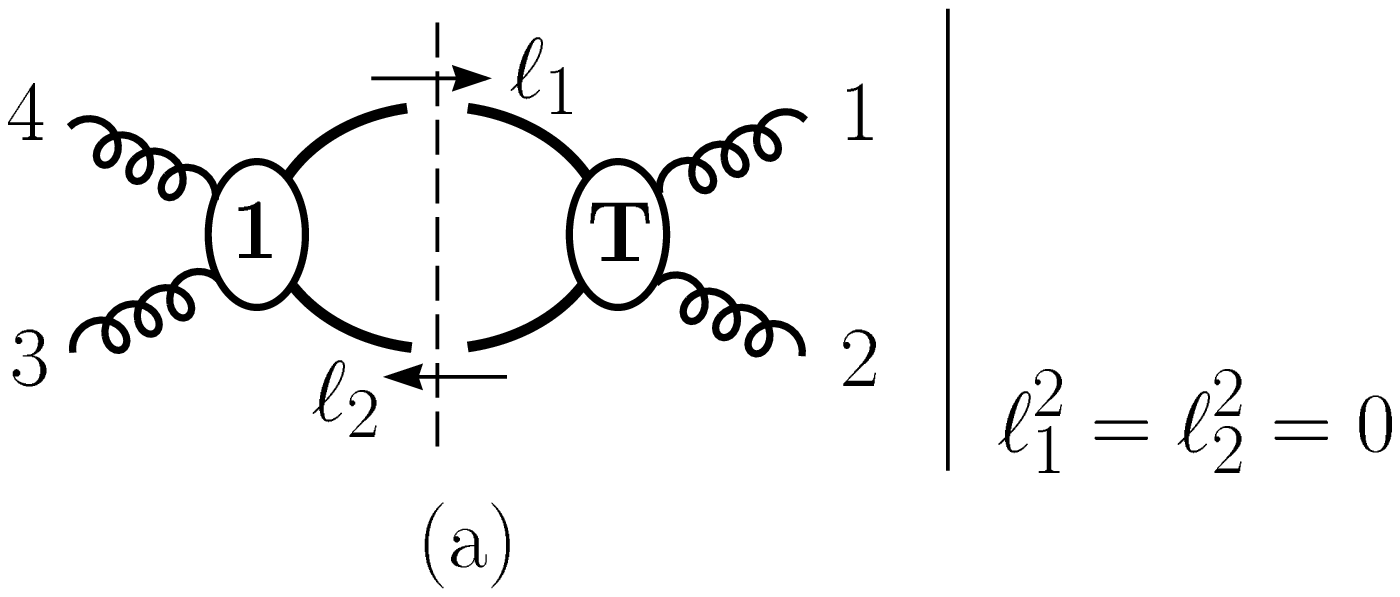,clip=,width=2.4in}
\hskip 1.5 cm
\epsfig{file=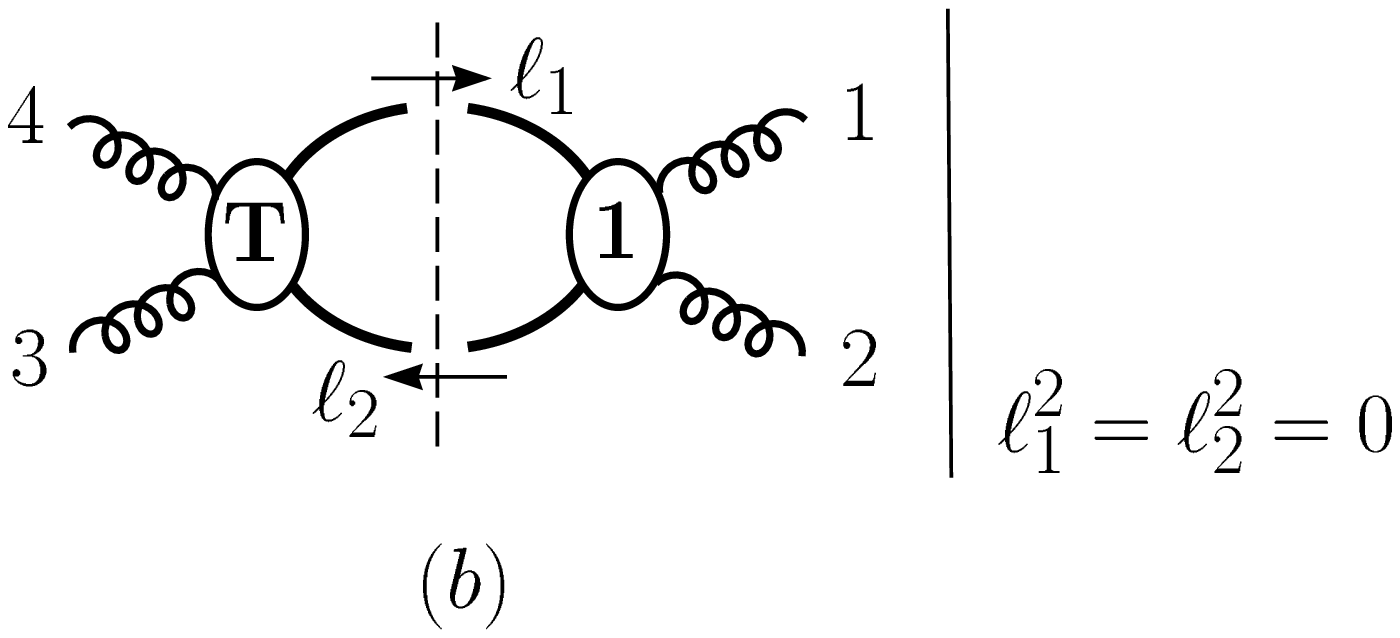,clip=,width=2.4in}
\end{center}
\vskip -.2 cm
\caption[]{
\label{TwoParticleFigure}
\small
The two-particle $s$-channel cut has two contributions: one with the 
four-point one-loop amplitude `1' to the left and the tree amplitude `T' to 
the right (a) and the other being the reverse (b).
}
\end{figure}

A convenient way to obtain amplitudes is from cuts of unrestricted
loop momentum integrals \cite{Review}.  In this way, one may
simultaneously construct the imaginary and real parts of the
amplitudes.  Consider for example, the cuts of the two-loop four-gluon
amplitude.  At two loops one must consider both two- and
three-particle cuts and in each channel multiple cuts can contribute.
For example, in \fig{TwoParticleFigure} the two $s$ channel
two-particle cuts are given.  By combining all cuts into a single
function, one obtains the full amplitude.

By computing cuts to higher orders in the dimensional regularization
parameter $\eps= (4-D)/2$ a complete reconstruction of a massless
amplitude is possible \cite{Unitarity,Review}. This follows from
dimensional analysis, since every term in an amplitude must have a
prefactor of powers of $(-s_{ij})^{-\eps}$ (where $s_{ij}= (k_i +
k_j)^2$), which necessarily has cuts.  For reasons of technical
simplicity, we use helicity amplitudes which implicitly take the cut
lines to have four-dimensional momenta.  In
ref.~\cite{TwoLoopSusyFour} we argued that for the two-loop four-point
$N=4$ amplitudes this does not introduce any errors at least through
$\Ord(\eps^0)$.  However, for QCD amplitudes one would need to compute
the contributions from the $(-2\eps)$-dimensional parts of of the loop
momenta, since they may interfere with poles in $\eps$ to produce
$\Ord(\eps^0)$ contributions.

\begin{figure}[ht]
\begin{center}
\vskip -.2 cm
\epsfig{file=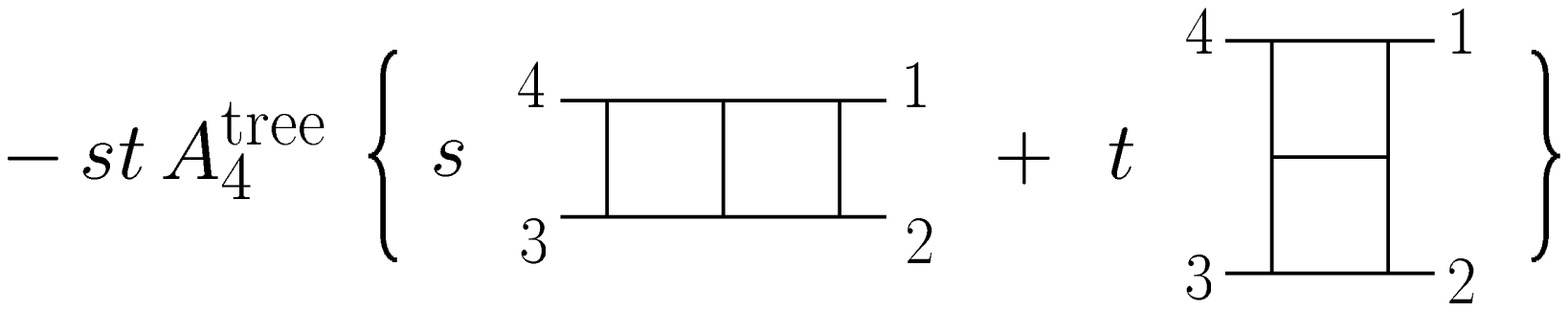,clip=,width=3.0 in}
\end{center}
\vskip -.2 cm
\caption[]{
\label{TwoLoopAmpl_ansFigure}
\small
The result for the leading color two-loop amplitude, corresponding to 
\eqn{LeadingColorResult}.  The diagrams correspond to scalar $\phi^3$
integrals.}
\end{figure}

Applying the cut-construction method to the leading color part of the
two-loop $N=4$ four-gluon partial amplitude, we obtain
$$
A_{4;1;1}^{\LC} = 
-s t \, A_4^\tree \left( s \, \I_4^{\P}(s,t)
+ t \, \I_4^{\P}(t,s) \right) \,.
\equn\label{LeadingColorResult}
$$
where $\I_4^{\P}(s,t)$ corresponds to the first $\phi^3$ scalar integral
in \fig{TwoLoopAmpl_ansFigure} and $\I_4^{\P}(t,s)$ to the second.

%

The computation of the subleading color pieces is similar to the 
leading color computation.  The complete two-loop four-gluon amplitudes
may be found in ref.~\cite{TwoLoopSusyFour}.

\begin{figure}[ht]
\begin{center}
\vskip -.4 cm
\epsfig{file=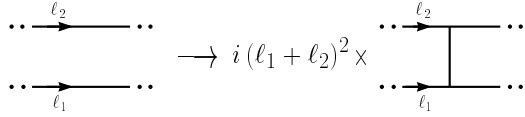,clip=,width=2.8in}
\end{center}
\vskip -.2 cm
\caption[]{
\label{AddLineFigure}
\small Starting from an $n$-loop integral function we may add an extra
line to obtain results for $(n+1)$-loop amplitudes.  The two-lines on
the left represent two lines buried in some $n$-loop integral.}
\end{figure}

Following the same cut construction procedure used for the two-loop
amplitudes, we have obtained a pattern for $n$-loop $N\! = \! 4$
four-gluon amplitudes: one takes each $n$-loop graph in the $n$-loop
amplitude and generates all the possible $(n+1)$-loop graphs by
inserting a new leg between each possible pair of internal
legs. Diagrams where triangle or bubble subgraphs are created should
not be included.  One also includes an additional factor of $i
(\ell_1+\ell_2)^2$ in the numerator, where $\ell_1$
and $\ell_2$ are the momenta flowing through each of the legs to which
the new line is joined, as depicted in \fig{AddLineFigure}.  Each
distinct $(n+1)$-loop graph should be counted once, even though they
can be generated in multiple ways. The $(n+1)$-loop amplitude is then
the sum of all distinct $(n+1)$-loop graphs multiplied by the tree
amplitude.

%

We have verified that this pattern is consistent with two-particle
cuts to all loop orders and with four-dimensional three-particle cuts,
up to five loops.  The two-particle cuts are not difficult to check
because the same algebra appears at any loop order.  The pattern remains a
conjecture because we have not verified the higher-particle cuts. The
same pattern also forms a conjecture for subleading in color contributions.

\begin{figure}[ht]
\begin{center}
\vskip -.3 cm
\epsfig{file=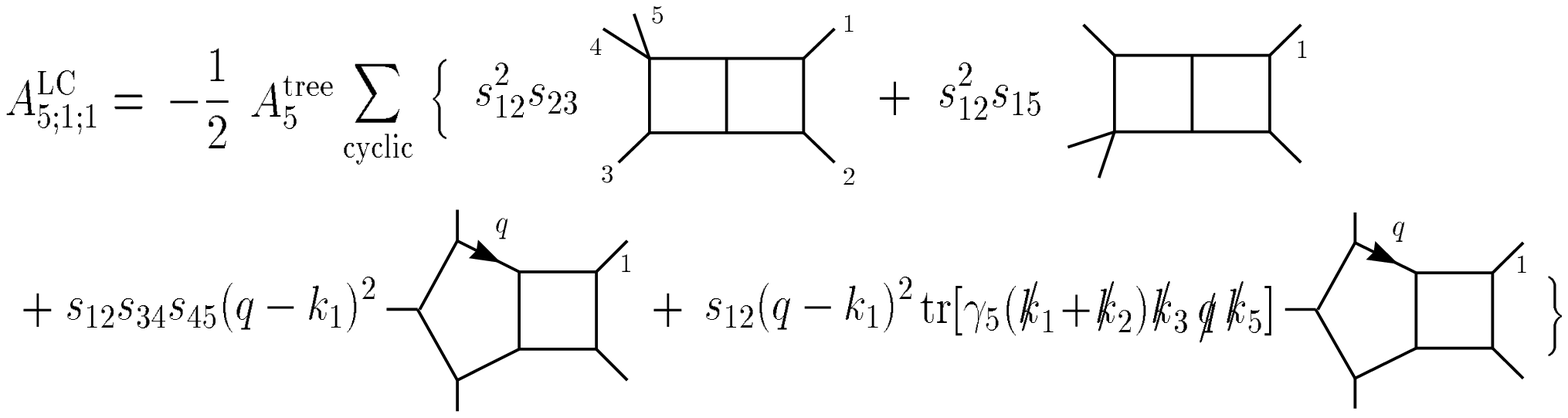,clip=,width=5in}
\end{center}
\vskip -.2 cm
\caption[]{
\label{FivePointFigure}
\small
A conjecture for the leading color parts of the two-loop five-gluon
amplitude. The arrow marks the line carrying loop momentum $q$.}
\end{figure}

These methods may also be applied to higher-point amplitudes.  In
\fig{FivePointFigure} we present a conjecture for the leading-color
part of five-point two-loop $N=4$ amplitudes.  The momentum $q$
appearing in the coefficients of the diagrams are loop momenta which
should be integrated over.  The complete consistency of this
conjecture has not been proven.

%

The equivalence between leading log contributions of the one-loop
$N=4$ amplitudes and FKL amplitudes \cite{FKL} has recently been
demonstrated~\cite{DelDuca}.  We comment that the conjecture for the
four-point multi-loop amplitudes presented here also has the expected
behavior.  Indeed, carrying out the leading log approximation
on the conjecture in \fig{AddLineFigure} yields,
$$
\eqalign{
A_4^{\rm tree} \left({s\over-t}\right)^{  \alpha(t)} \hskip -.5 cm 
, \hskip 1. cm 
\alpha(t) = 2\, g^2 \,N_c \,{1\over \eps}\, {1\over (4\pi)^{2-\eps}} 
{\Gamma(1+\eps) \Gamma^2(1-\eps) \over \Gamma(1-2\eps)} \, 
\Bigl({\mu^2 \over -t}\Bigr)^\eps \,. 
}
\equn\label{LeadingLogAllLoop}
$$

\section*{Conclusions}

In this talk we outlined the initial steps that have been taken in
applying the cut construction technique reviewed in ref.~\cite{Review}
to two-loop amplitudes.  We expect that the cutting methods discussed
here will be applied to two-loop QCD amplitudes.  However, before
physics can be extracted one would need numerically stable expressions
for loop integrals and for the infrared divergent corners of phase
space.

\vskip .2 cm 

We thank L. Dixon and D.A.~Kosower for many helpful discussions. We
also thank V. Del Duca and C. Schmidt for notes on the relationship of
$N=4$ and FKL amplitudes.

\end{document}